\documentclass[11pt]{article}
\usepackage{graphicx}
\usepackage[margin=1in]{geometry}
\usepackage{natbib}
\usepackage[bf,font={small,sl}]{caption}
\usepackage{parskip}
\usepackage{bm,amsmath,amsfonts,tikz}
\usetikzlibrary {shapes ,arrows , positioning}
\usepackage{todonotes}
\usepackage{url}
\usepackage[affil-it]{authblk}

\newcommand{\Y}{\bm{Y}}

\title{\textbf{Scale dependence in hidden Markov models\\ for animal movement}}
\author{Théo Michelot$^1$\footnote{Email: \texttt{theo.michelot@dal.ca}}, Emma Storey$^2$}
\affil{$^1$Dalhousie University, Canada\\ $^2$Centre for Environment, Fisheries and Aquaculture Science, UK}
\date{}

\begin{document}
\maketitle

\begin{abstract}\it\noindent
    Hidden Markov models (HMMs) have been used increasingly to understand how movement patterns of animals arise from behavioural states. An animal is assumed to transition between behavioural states through time, as described by transition probabilities. Within each state, the movement typically follows a discrete-time random walk, where steps between successive observed locations are described in terms of step lengths (related to speed) and turning angles (related to tortuosity). HMMs are discrete-time models, and most of their outputs strongly depend on the temporal resolution of data. We compile known theoretical results about scale dependence in Markov chains and correlated random walks, which are the most common components of HMMs for animal movement. We also illustrate this phenomenon using simulations covering a wide range of biological scenarios. The scale dependence affects not only all model parameters, i.e., the transition probabilities and the movement parameters within each behavioural state, but also the overall classification of movement patterns into states. This highlights the importance of carefully considering the time resolution when drawing conclusions from the results of analysis. In addition, scale dependence generally precludes the analysis of tracking data collected at irregular time intervals, and the comparison (or combination) of data sets with different sampling rates. HMMs remain a valuable tool to answer questions about animal movement and behaviour, as long as these limitations are well understood.
\end{abstract}

\section{Introduction}

Statistical models for animal tracking data are used to understand the behaviour of animals, their relationships to their environment, and the emergence of their spatial distributions \citep{hooten2017}. One important modelling choice is the time formulation: should movement be modelled in discrete time as a sequence of steps, or in continuous time \citep{mcclintock2014}? Discrete-time models based on random walks are often easier to work with mathematically, but they are scale-dependent: their interpretation is tied to the time interval of observation. This is a crucial limitation, and often precludes comparing or combining studies with different sampling frequencies. 

There has been extensive research to understand how properties of discrete-time models scale with the length of the time interval, based both on theoretical work and simulations. For example, the seminal paper of \cite{kareiva1983} derived a scaling formula for the mean squared displacement of correlated random walks. \cite{benhamou2004} studied the effect of time interval on the observed sinuosity of animal paths.  \cite{codling2005} used simulations to investigate the effect of sampling rate on various parameters of biased and correlated random walks (including speed and directional persistence). \cite{pepin2004}, \cite{rowcliffe2012} and \cite{noonan2019} observed that the distance travelled by animals can be severely underestimated based on tracking data, if one assumes that animals follow straight segments between observations.

In this short note, we investigate the scale dependence of hidden Markov models (HMMs), a widely-used method to study the behavioural process that underlies animal movement \citep{morales2004, langrock2012}. Because HMMs combine Markov chains and random walks, this work is broadly a compilation of known results about those component models, illustrated with simulations. We place these results within the context of HMMs as they are commonly used in movement ecology, and discuss some of the implications.

\section{Overview of hidden Markov models for animal movement}
\label{sec:hmm}

\subsection{Model formulation}

A hidden Markov models (HMM) is a time series model comprising an observed process $(\Y_1,\dots,\Y_T)$ and an underlying unobserved state process $(S_1,\dots,S_T)$. In the context of animal movement, the observed variables are typically metrics derived from telemetry locations $\bm{x}_1, \bm{x}_2, \dots$ collected at regular time intervals. Following the most common formulation, we choose $\Y_t = (l_t, \phi_t)$, where the step length $l_t$ is the straight-line distance between $\bm{x}_{t}$ and $\bm{x}_{t+1}$, and the turning angle $\phi_t$ is the directional change between $\bm{x}_{t-1} \rightarrow \bm{x}_t$ and $\bm{x}_{t} \rightarrow \bm{x}_{t+1}$ \citep{morales2004, patterson2009, langrock2012}. The observation $\Y_t$ is generated by one of $N$ distributions, as determined by the underlying state $S_t \in \{ 1, \dots, N\}$ at time $t$. In animal movement analyses, the hidden state is often interpreted as a proxy for the behavioural state of the tracked animal, such as `foraging' or `travelling' \citep{morales2004}. When the observation process is specified using step lengths and turning angles, this HMM formulation can be viewed as a state-switching version of the correlated random walk \citep{kareiva1983, bovet1988, morales2004}.

The state process is modelled as a first-order Markov chain, i.e., the probability of a state being active at time $t$ is conditionally independent of the history of the process until time $t-2$, given the state at time $t-1$: $\Pr(S_t \mid S_{t-1}, \dots, S_1) = \Pr(S_t \mid S_{t-1})$. Additionally, observations are assumed to be conditionally independent from past observations and states, given the current state: $p(\Y_t \mid \Y_{t-1}, \dots ,\Y_1, S_t, \dots, S_1) = p(\Y_t \mid S_t)$. When modelling animal movement data, it is convenient to make the further assumption that the observed variables (i.e., step lengths and turning angles) are conditionally independent given the current state: $p(\Y_t \mid S_t = i) = p(l_t, \phi_t \mid S_t = i) = p(l_t \mid S_t = i) p(\phi_t \mid S_t = i)$ for $i = 1, \dots, N$ \citep{morales2004}.

The state process is parameterised in terms of an initial distribution and transition probabilities. The initial distribution $\bm\delta^{(1)}$ is a vector of the probabilities of being in each state at the first time step, $\delta^{(1)}_i = \Pr(S_1 = i)$ for $i = 1, \dots, N$. The state transition probabilities describe the dynamics of the hidden process, and they are often written as an $N \times N$ transition probability matrix
\begin{equation*}
    \bm{\Gamma} =
    \begin{bmatrix}
    \gamma_{11} & \gamma_{12} & \dots & \gamma_{1N} \\
    \gamma_{21} & \gamma_{22} & \dots & \gamma_{2N} \\
    \vdots      & \vdots     &\ddots& \vdots \\
    \gamma_{N1} & \gamma_{N2} & \dots & \gamma_{NN} \\
    \end{bmatrix}   
\end{equation*}
where, for $i, j \in \{ 1, 2, \dots, N\}$, $\gamma_{ij} = \Pr(S_t = j \mid S_{t-1} = i)$ denotes the probability of switching from state $i$ to state $j$ over one time interval of observation. Each row of the transition probability matrix is constrained to sum to one, i.e., $\sum_{j=1}^N \gamma_{ij} = 1$ for all $i = 1, \dots, N$. 

A common goal of HMM analyses of animal movement is to explore drivers of behaviour. As such, it is often of interest to incorporate time-varying covariates within the hidden process, for example to capture the effects of environmental conditions on an animal's behaviour \citep{morales2004, patterson2009}. The model can be extended to allow transition probabilities to vary over time, such that $\bm{\Gamma}^{(t)} = (\gamma_{ij}^{(t)})$. The transition probabilities can then be linked to $p$ covariates $x_{1t}, \dots, x_{pt}$ via the multinomial logit link,
\begin{align}
    \gamma_{ij}^{(t)} = 
    \Pr(S_t = j \mid S_{t-1} = i) = 
    \frac{\exp(\eta_{ij}^{(t)})}{\sum_{k=1}^N \exp(\eta_{ik}^{(t)})}, \label{eqn:multinom} \\
    \text{where}\quad
    \eta_{ij}^{(t)} = 
    \begin{cases}
        \beta_0^{(ij)} + \beta_1^{(ij)} x_{1t} + \dots \beta_p^{(ij)} x_{pt} & \text{if } i \neq j, \\
        0 & \text{if } i = j.
    \end{cases}\nonumber
\end{align}
This form ensures that transition probabilities are between 0 and 1, and that each row of $\bm\Gamma^{(t)}$ sums to one.

For the observation model, each observed variable is typically modelled by a distribution with state-dependent parameters, i.e., $Y_t \mid S_t = j \sim \mathcal{D}(\bm\theta^{(j)})$, where $\mathcal{D}$ could for example represent the normal or gamma distribution, and $\bm\theta^{(j)}$ is the vector of parameters in state $j$. In this paper, we model step length using the gamma distribution, and turning angle with the von Mises distribution, which are both common choices in HMM analyses \citep{morales2004, michelot2016}. The observation model can therefore be written as
\begin{equation*}
    \begin{cases}
        l_t \mid S_t = j \sim \text{gamma}(\mu_j, \sigma_j) \\
        \phi_t \mid S_t = j \sim \text{von Mises}(\lambda_j, \kappa_j)
    \end{cases}
\end{equation*}
where $\mu_j$ is the mean step length, $\sigma_j$ is the standard deviation of step lengths, $\lambda_j$ is the mean turning angle, and $\kappa_j$ is the concentration of turning angles. We use the parameterisation of the gamma distribution in terms of a mean and standard deviation, rather than shape and scale, because we are interested in the relationship between mean step length and temporal scale. Although we focus on these particular distributions for illustration, the issue of scale dependence would apply similarly to any other distribution.

\subsection{Inference}

We assume that locations of an animal are collected at regular time intervals with negligible measurement error, and we use them to derive step lengths $l_1, l_2, \dots$, and turning angles $\phi_1, \phi_2, \dots$. From these data, statistical inference can be used to estimate all model parameters, i.e., the initial distribution and transition probabilities of the state process, and the state-dependent observation parameters. This can be done using maximum likelihood estimation based on algorithms that efficiently account for all possible values of the unobserved state process \citep[e.g., the forward algorithm;][]{langrock2012, zucchini2017}. After model fitting, the most likely state sequence can also be computed using the Viterbi algorithm, to classify the observed time series into states. The R packages moveHMM and momentuHMM provide convenient functions to analyse animal movement data with hidden Markov models \citep{michelot2016, mcclintock2018}.

\section{Scale-dependent model outputs}

\begin{figure}[htbp]
    \centering
    \includegraphics[width=0.7\textwidth]{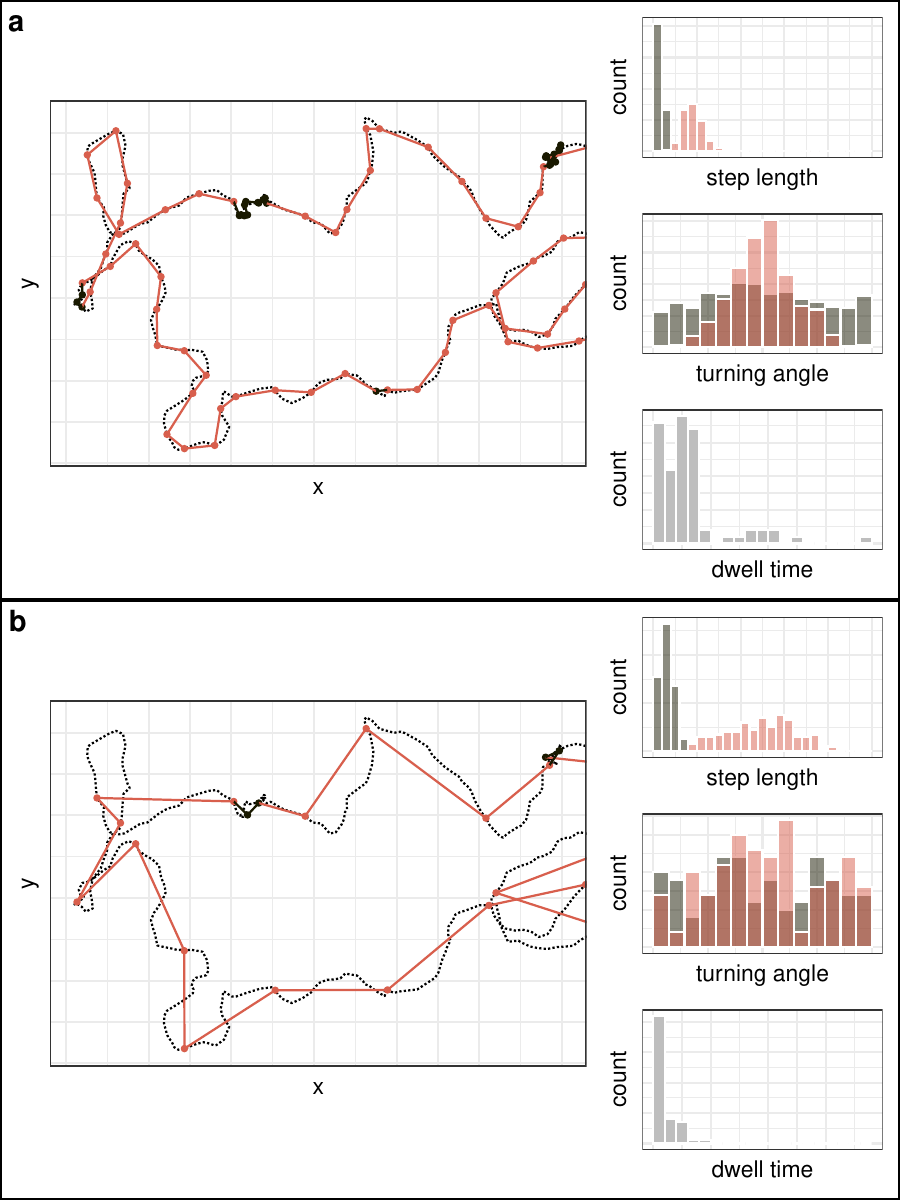}
    \caption{Overview of scale dependence in hidden Markov models (HMMs). Panels (a) and (b) show data sets derived at different time resolutions from the same simulated trajectory, and HMM outputs. Left: true continuous track (dotted line) and observed locations (dots) coloured by the most likely state sequence obtained from a 2-state HMM. Right: histograms of movement metrics grouped by estimated states, and histograms of dwell times (number of time steps between state transitions). As the time interval gets longer, step lengths tend to become larger, turning angles more spread out, and dwell times shorter.}
    \label{fig:overview}
\end{figure}

All the outputs of a typical HMM analysis are dependent on the temporal resolution of the data, including the movement parameters within each state, the transition probabilities describing the dynamics of the state process, and the clustering of observations into states. This general idea is illustrated with an example in Figure \ref{fig:overview}. In that figure, two data sets with different time resolutions are shown, for the same underlying continuous animal track. It is clear that observed step lengths and turning angles change with the time interval of observation; for example, step lengths tend to be longer at coarser time resolutions. However, the exact scaling laws of movement and behavioural metrics are not always known.

In this section, we use simulations to illustrate the effect of changing the temporal resolution of tracking data on estimated HMM parameters and other outputs. All simulations were implemented using the R package moveHMM version 1.9 \citep{michelot2016}, and simulation code is available online at \url{https://github.com/TheoMichelot/HMM_scale_dependence}. We consider two aspects of scale dependence in HMMs: in the movement patterns within each state, and in behavioural dynamics.

\subsection{Scale-dependent movement patterns within each state}

The observed movement patterns change with the time interval of observation, in particular when summarised into step lengths and turning angles. The movement model within each state can therefore only be interpreted in relation to that interval; this is not specific to HMMs, and it applies to all discrete-time random walk models \citep{kareiva1983, turchin1998, codling2008}.

\subsubsection{Step length parameters}
\label{sec:step-par}

The relationship between distance travelled and time interval has been studied extensively in the context of correlated random walks \citep{kareiva1983, turchin1998, codling2008}. The focus of that work has usually been the mean squared displacement (or mean squared distance; MSD), i.e., the expected squared straight-line distance between the location of the animal at time $0$ and its location $t$ time units later. \cite{kareiva1983} derived a formula for the MSD as a function of $t$ in a correlated random walk, and showed that it depends on the strength of correlation (i.e., the amount of directional persistence). For non-correlated movement the MSD is proportional to $t$, whereas for very highly correlated movement it is proportional to $t^2$; most animals' movement is somewhere between these two extremes. The MSD is closely related to the step length modelled in an HMM: the MSD after $t$ time units can be thought of as the mean squared step length measured when locations are observed at intervals of $t$ time units. One implication of the results of \cite{kareiva1983} is that the step length usually does not scale linearly with the time interval, because animals do not move in a completely straight line. 

\begin{figure}[htbp]
    \centering
    \includegraphics[width=\textwidth]{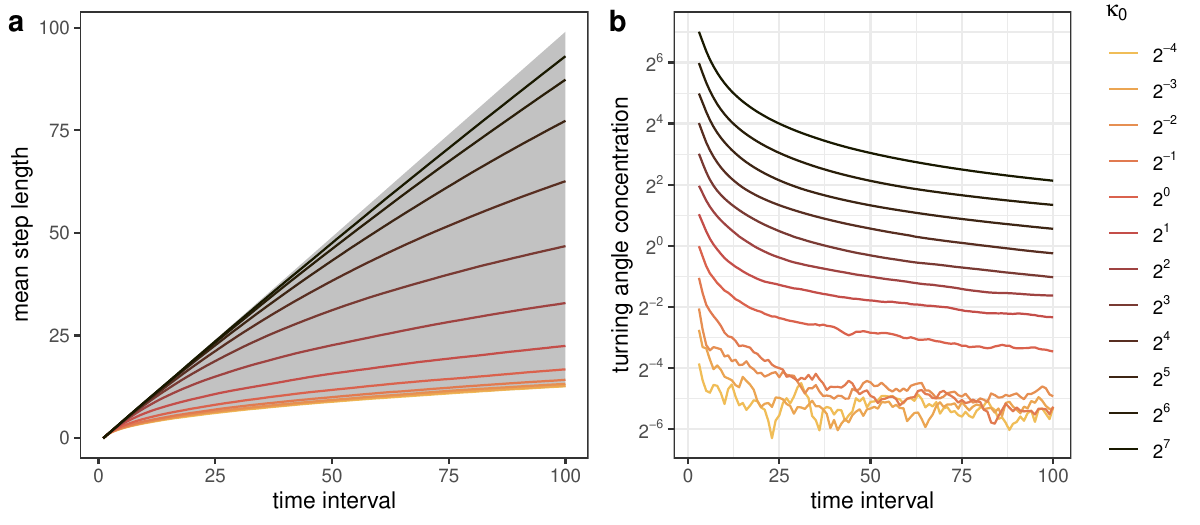}
    \caption{Results of simulation on scale dependence of correlated random walk parameters. Mean step length (a) and turning angle concentration on the log scale (b), plotted against time interval of observation. Both simulations were repeated for different levels of movement persistence, as measured by simulation concentration $\kappa_0$. In (a), the theoretical upper and lower bounds are shown as a grey ribbon. Darker lines correspond to simulated trajectories with stronger directional persistence.} 
    \label{fig:move-par}
\end{figure}

The relationship between step length and time interval is illustrated using simulations in Figure \ref{fig:move-par}(a). We simulated from correlated random walks with different amounts of correlation, i.e., different turning angle concentrations between $\kappa_0 = 2^{-4} = 0.0625$ and $\kappa_0 = 2^7 = 128$. For each $\kappa_0$, we simulated 100 observations, and then measured the distances to each point in the track to the start point, analogous to step lengths over different time intervals. We repeated this process 10,000 times, and measured the mean step length for each time interval. The relationship is clearly non-linear in general, although it approaches a straight line when persistence in direction is extremely high. 

If step lengths over one time interval follow a gamma distribution with mean $\mu_0$ and standard deviation $\sigma_0$, theoretical results suggest that the scaling law for the mean step length over $n$ time intervals is approximately bounded between $\sqrt{\pi(\mu_0^2+\sigma_0^2)}/2 \cdot \sqrt{n}$ (in the absence of directional persistence) and $\mu_0 \cdot n$ (for perfect directional persistence). These bounds are derived in Appendix \ref{app:steplength} and shown in Figure \ref{fig:move-par}(a), and match the simulation results. 

To reduce the dependence of step length on the time interval of observation, one apparent solution is to compute ``movement rates'', i.e., step lengths divided by the time interval. However, this is generally not a good measure of an animal's movement speed, because it assumes that it moves along the observed segments of the track, and it therefore does not adequately account for scale dependence \citep{postlethwaite2013, noonan2019}. It is preferable to work with the original step lengths instead, and explicitly state that all inferences should be interpreted at the time interval of observation.

\subsubsection{Turning angles in correlated random walk}

Turning angle is a useful variable to measure persistence (or, conversely, sinuosity) in an animal's movement, and it is used to formulate correlated random walks \citep{kareiva1983, bovet1988}. Likewise, it is often included as a variable of interest in HMMs, and modelled with state-dependent circular distributions \citep{morales2004, langrock2012, michelot2016}. The von Mises and wrapped Cauchy distributions, often used for this purpose, both have two parameters: mean, and angular concentration. In most applications, the mean turning angle should be close to zero, representing directional persistence, and concentration is the main parameter used to learn about the animal's behaviour. The turning angle concentration parameter measures how peaked the distribution is around its mean, and it is inversely related to variance. 

The sinuosity of an observed movement path depends on the time resolution: an animal's path will always display strong directional persistence at a very fine time scale \citep[``because, put simply, all life has a front and rear end'';][]{pyke2015}, but this persistence decays at coarser resolutions \citep{codling2005, postlethwaite2013}. We investigated the dependence of turning angle distribution on the time interval of observation using simulations, and the results are shown in Figure \ref{fig:move-par}(b). We simulated correlated random walks at a fine time resolution using the same grid of turning angle concentrations used in Section \ref{sec:step-par} (from $\kappa_0 = 2^{-4}$ to $\kappa_0 = 2^7$), over 100 time steps. We then measured the turning angles that would be obtained at each time resolution between 1 and 100 time units. We repeated this procedure 10,000 times, and computed the turning angle concentration at each time resolution. Figure \ref{fig:move-par}(b) shows that the turning angle concentration tends to decrease with time interval, consistent with the intuition that movement persistence will appear to decrease over long intervals. For very coarse data, the concentration approaches zero, which corresponds to a uniform distribution over $(-\pi, \pi]$ (i.e., an uncorrelated random walk).

\subsubsection{Turning angles in biased random walk}

When an animal moves around a centre of attraction, we might also expect the mean of the turning angle distribution to depend on the sampling interval. Indeed, at coarse temporal resolutions, this type of movement yields many apparent reversals in direction, and the observed turning angles can have a mean close to $\pi$. As mentioned above, all animals display some directional persistence at a fine resolution, and a mean turning angle of $\pi$ should be viewed as an artefact of the discrete-time observation scheme. 

We simulated tracks from a biased correlated random walk, where the movement direction at each time step is a mixture of persistence and bias towards a centre of attraction, as described by \cite{duchesne2015}. We downsampled the tracks to observation intervals $\Delta = 1, 2, \dots, 100$ time units and, for each interval, we fitted a von Mises distribution to the observed turning angles (mean and concentration). We derived the cosine of the estimated mean turning angle, which is a convenient measure of directional persistence: a cosine of 1 corresponds to a mean turning angle of zero (persistence in direction), and a cosine of -1 corresponds to a mean turning angle of $\pi$ or $-\pi$ (reversal in direction). We used the cosine rather than the mean turning angle itself because, when there is reversal in direction, the mean can arbitrarily be close to $-\pi$ or $\pi$. We ran this procedure 1000 times, and computed averages of the turning angle parameters for each time interval of observation. We repeated the simulation for different mixtures of persistence and bias: high bias, low bias, and no bias (i.e., CRW).

\begin{figure}[htbp]
    \centering
    \includegraphics[width=\textwidth]{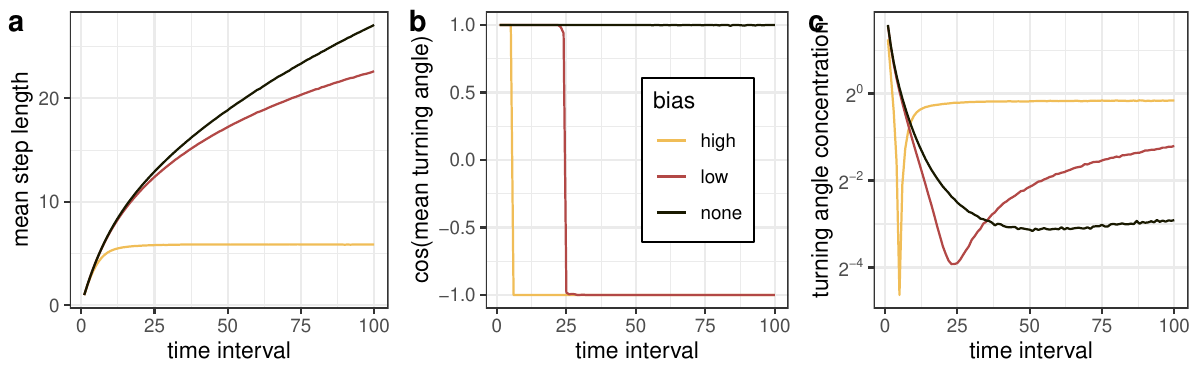}
    \caption{Results of biased correlated random walk simulation: (a) mean step length, (b) cosine of mean turning angle, and (c) turning angle concentration, as functions of time interval of observation. The cosine of the mean turning angle changes from $1$ (persistence in direction) to $-1$ (apparent reversal in direction) when the sampling resolution gets coarser. The colours show three scenarios with different amounts of bias.}
    \label{fig:bcrw}
\end{figure}

Estimated turning angle parameters are shown in Figure \ref{fig:bcrw}. The turning angle distribution was centred on zero at fine time resolutions (cosine = 1) for all three scenarios, indicating directional persistence. In the presence of bias, the observed mean suddenly changed to $\pi$ (or $-\pi$) for intervals longer than some threshold (cosine = $-1$). This suggests that reversals in direction dominate the distribution of turning angles at coarse resolutions, and the threshold is smaller when there is strong attraction to the centre. When there was no bias, the mean turning angle remained around zero at any time resolution, and the concentration parameter decreased to zero (similar to results from the previous section). However, in the biased random walks, the turning angle concentration did not consistently decrease as the time interval increases; it decreased at first, when the mean turning angle was zero, but it started increasing when the turning angle was estimated to $\pi$. This is because reversals in direction become more and more frequent at coarser time resolutions, resulting in a higher peak of turning angles around $\pi$.

\subsection{Scale-dependent state dynamics}

\subsubsection{Transition probabilities}
\label{sec:trprobs}

The unobserved state process of an HMM is specified in terms of the probabilities of a transition between any two states over one time interval, and these parameters can in turn be used to derive expected dwell times (number of intervals between transitions) and activity budgets (long-term proportion of time in each state). It is well known in the literature on Markov chains that the transition probabilities scale with the time interval.

The $n$-step transition probabilities of a Markov chain are defined as $\gamma_{ij}^{(n)} = \Pr(S_{t+n} = j \mid S_t = i)$. These are closely related to the (one-step) transition probabilities of the process: the $n$-step transition probability matrix is equal to $\bm{\Gamma}^n$ \citep[Section 4.2 of][]{ross2019}. In the 2-state case, we find the following relationship between the transition probabilities and the time interval:
\begin{equation}
\label{eqn:nstep}
    \gamma_{ij}^{(n)} = \frac{\gamma_{ij}^{(0)}}{\gamma_{12}^{(0)} + \gamma_{21}^{(0)}} \left[ 1 - (1 - \gamma_{12}^{(0)} - \gamma_{21}^{(0)})^n \right]
\end{equation}
and this is shown in Appendix \ref{app:tpm}. We illustrate this below using simulations. 

\begin{figure}[htbp]
    \centering
    \includegraphics[width=0.48\textwidth]{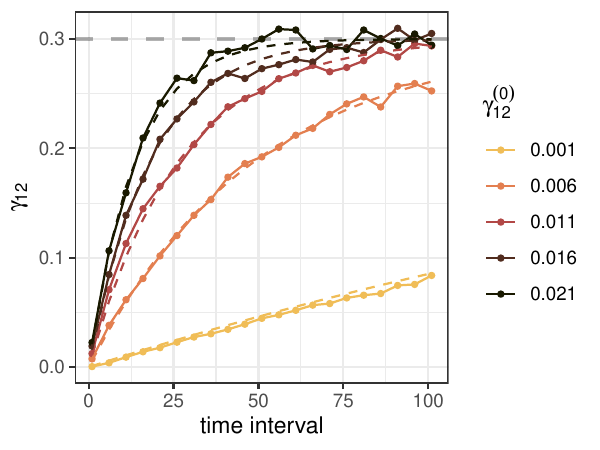}
    \includegraphics[width=0.48\textwidth]{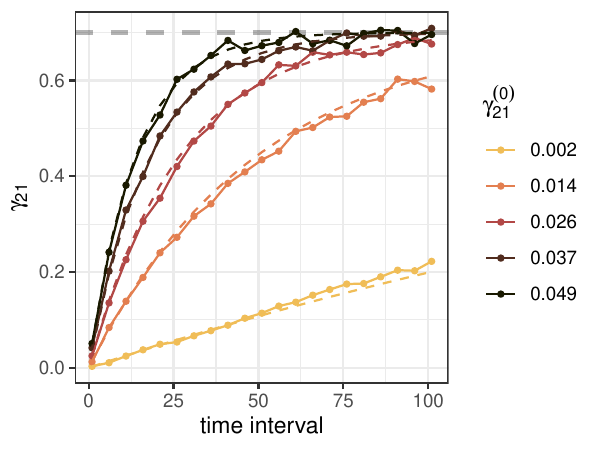}
    \caption{Results of simulations on scale dependence of transition probabilities. The solid lines show the estimated transition probabilities $\gamma_{12}$ and $\gamma_{21}$ as functions of the time interval of observation, for different simulation scenarios (i.e., different transition probabilities $\gamma^{(0)}_{12}$ and $\gamma^{(0)}_{21}$ used to simulate at a fine time resolution). The darker lines correspond to scenarios where state switches are more frequent. In all simulation scenarios, the stationary distribution of the state process is $(0.7, 0.3)$, and this is shown by horizontal dashed lines.}
    \label{fig:tpm}
\end{figure}

We simulated movement trajectories from 2-state HMMs with different transition probabilities, all with the same stationary distribution $\pi = (0.7, 0.3)$. For each set of transition probabilities, movement tracks were simulated at a fine time resolution (i.e. $\Delta$ = 1 time unit), and 2-state HMMs were fitted to down-sampled tracks over regular intervals of $\Delta = 1, 6, 11, \dots, 101$ time units. Movement tracks were generated such that there were 5000 observations at the coarsest time resolution, to ensure that longer time intervals were not subject to bias arising from small sample sizes. Figure \ref{fig:tpm} shows the estimated transition probabilities against the time interval. As the temporal resolution gets coarser, the transition probabilities between states 1 and 2 tended to increase, consistent with the intuition that as the time interval increases relative to the scale of behaviour, the persistence of the behaviour will appear to decay. The theoretical relationships between $n$-step transition probabilities and time interval are also displayed in the figure, and they closely match the simulation results.

In Figure \ref{fig:tpm}, the $n$-step transition probabilities plateau as the time interval of observation becomes large. This asymptotic behaviour reflects the long-term limiting distribution of the Markov chain: as $t \to \infty$, the transition probabilities converge to the stationary distribution. Convergence of the $n$-step transition probabilities to the stationary distribution occurs if a Markov chain is irreducible (i.e., it is possible to travel between any two states) and aperiodic (i.e., states do not occur at a regular number of time steps); these conditions are typically met in HMMs of animal movement. For a 2-state HMM, this can be seen in Equation \ref{eqn:nstep} by taking the limit $n \to \infty$,
\begin{equation*}
    \pi = 
    \left(
    \frac{\gamma_{21}^{(0)}}{\gamma_{12}^{(0)} + \gamma_{21}^{(0)}},\ 
    \frac{\gamma_{12}^{(0)}}{\gamma_{12}^{(0)} + \gamma_{21}^{(0)}}
    \right).
\end{equation*}

Intuitively, if two observed locations of an animal are separated by a very long period, the probability that the animal ends the step in a given behavioural state does not depend on its state at the start of the step. This probability is just the long-term proportion of time spent in that state.

\subsubsection{Covariate effects}

HMMs are commonly used to understand the drivers of animals' behavioural changes, by including covariates on the transition probabilities \citep{patterson2009}. The output of interest is then the regression coefficients linking the transition probabilities to the covariates, or, more generally, the shape of the relationship \citep[which could be non-linear;][]{michelot2025}. The scale dependence of the transition probabilities implies that those quantities will also depend on the time interval.

We simulated tracking data similarly to the procedure described in Section \ref{sec:trprobs} but, this time, we included the effect of a covariate on the transition probability from state 1 to state 2. We used the multinomial logistic formulation given in Equation \ref{eqn:multinom}, and considered three different slope coefficients for the relationship: $\beta_1^{(12)} \in \{ -5, 2, 8\}$. We then downsampled the simulated data to $\Delta = 1, 6, 11, \dots, 101$ time units, and we fitted HMMs to the thinned data to see if the effect of the covariate could be recovered. We repeated this process 100 times, and calculated summary statistics of the parameter estimates over a grid of time resolutions. 

\begin{figure}[htbp]
    \centering
    \includegraphics[width=0.7\textwidth]{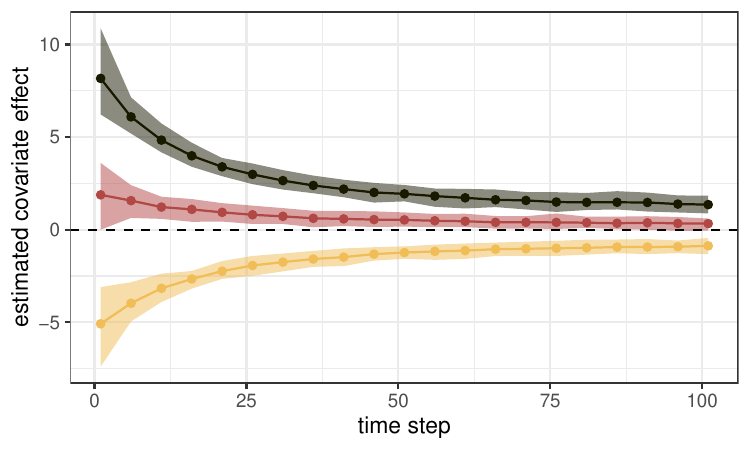}
    \caption{Results of simulation on scale dependence of estimated covariate effects on transition probability. The dots are mean point estimates across 100 simulations, for the regression coefficient $\beta_1^{(12)}$ linking the transition probability $\gamma_{12}$ to the covariate, based on data downsampled to different resolutions between 1 and 101 time units. The three lines correspond to different values of the simulation parameter. The shaded bands are 95\% quantile regions for the 100 simulations.}
    \label{fig:tpm-cov}
\end{figure}

Results shown in Figure \ref{fig:tpm-cov} suggest that the estimated covariate effects tended to decrease as the time interval between observations increased. Interestingly, at very coarse resolutions, the estimates plateaued to values that were still clearly different from zero. That is, the model was still able to identify whether the slope of the effect was positive or negative, though the amplitude varied widely with time resolution. This might be because dependence of the transition probabilities on covariates implies a similar dependence for the stationary distribution of the Markov chain. Then, if the estimated probabilities converge to the stationary distribution (as illustrated in Section \ref{sec:trprobs}), a relationship can still be found even for very sparse data.

\subsubsection{State classification}

In Section \ref{sec:trprobs}, we showed that the transition probabilities of a discrete-time Markov chain (used to specify an HMM) have a complex non-linear relationship with the time interval. One additional complication is that the \emph{definition} of the HMM states, or what behaviours they correspond to, is also susceptible to change for different sampling rates. There are therefore at least two sources of scale dependence in the dynamics of the state process: the parameter interpretation changes with the time resolution, and each HMM state might not correspond to the same behaviours at different sampling rates. There is generally no guarantee that HMM states will match biologically-relevant behavioural states \citep{patterson2017}, and this challenge is compounded with dependence on the sampling scheme.

We simulated data from a hierarchical HMM, as described by \cite{leos2017}, as it describes a realistic situation where we expect the state definition to change with sampling rate. We simulated two nested Markov chains, representing an animal's behavioural processes at two different scales. The coarse-scale process was a 2-state Markov chain with time intervals of 100 time units, with transition probabilities $\gamma_{12} = \gamma_{21} = 0.9$. The fine-scale process was a 2-state Markov chain at time intervals of 1 time unit, and its transition probability matrix depended on the currently-active coarse-scale state. Those were chosen so that the coarse-scale states differed in the proportion of time spent in each fine-scale state. We then simulated a movement track at the fine resolution (i.e., 1 time unit) based on the fine-scale simulated state sequence, where the two states were characterised by different movement parameters. To investigate the effect of sampling scheme, we fitted 2-state HMMs to downsampled tracks with various intervals $\Delta = 1, 2, \dots, 12$ time units. 

\begin{figure}[htbp]
    \centering
    \includegraphics[width=\textwidth]{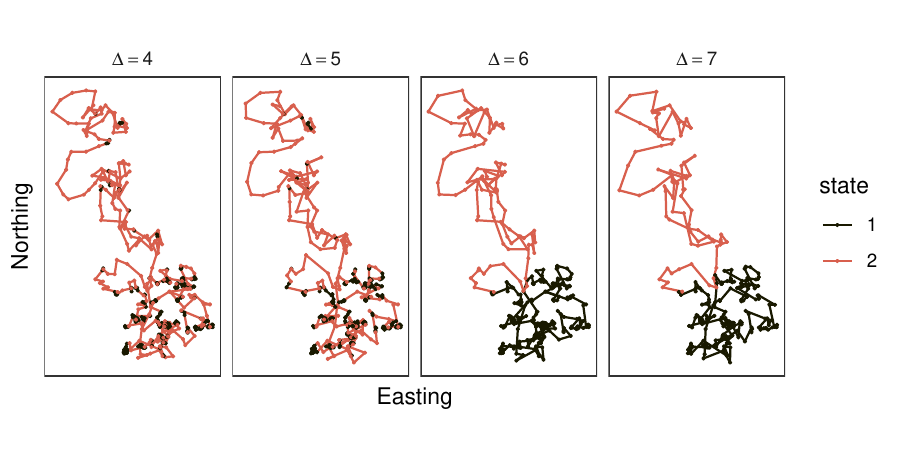}
    \caption{Results of simulation on scale dependence of state classification. The different panels show a movement track downsampled to different time resolutions ($\Delta \in \{4, 5, 6, 7\}$), to each of which an HMM has been fitted. The two colours correspond to the most likely state sequence obtained from each HMM. A qualitative change in the state classification occurs between $\Delta = 5$ and $\Delta = 6$.}
    \label{fig:multiscale}
\end{figure}

Figure \ref{fig:multiscale} shows the most likely state sequence computed from the Viterbi algorithm for a subset of the data at different time resolutions. The state classification remained relatively unchanged between $\Delta = 1$ and $\Delta = 5$, but it suddenly changed for $\Delta = 6$ and longer intervals. This is because, for $1 \leq \Delta \leq 5$, the states identified by the HMM matched the fine-scale states from the simulation, whereas, for $\Delta \geq 6$, the HMM states corresponded to the coarse-scale simulated states.

The simulations demonstrate that the state classification can change abruptly at different time resolutions of observations, in cases where the animal follows nested behaviours at different time scales. In practice, this is likely to apply to many study systems, because the true behavioural processes of animals are complex. Even if one could in principle find scaling laws for all movement and behavioural parameters, this would therefore still not guarantee that analyses performed at different time intervals are describing the same behavioural states.

\section{Conclusion}

This note does not intend to discourage practitioners from using HMMs and other discrete-time models, but to emphasise the dependence of model outputs on a particular time interval. We summarise the main points below.
\begin{enumerate}
    \item Apparent speed and mean step length decrease at longer time intervals, and the relationship depends on the amount of directional persistence.
    \item Apparent directional persistence decreases at longer time intervals.
    \item Home range behaviour leads to apparent reversals in direction at long time intervals.
    \item Estimated transition probabilities converge to the stationary distribution of the Markov chain at long time intervals.
    \item Estimated effects of covariates on transition probabilities tend to decrease as time interval increases.
    \item The segmentation of tracks into states can change qualitatively (and unpredictably) as time resolution changes, because what ``behaviours'' can be detected depends on sampling rate.
\end{enumerate}

These observations suggest that caution is needed to interpret results from discrete-time models, and provide strong motivation for further development of continuous-time methods that tend to circumvent such problems \citep[e.g.,][]{johnson2008, blackwell2016, noonan2019}.

\bibliographystyle{apalike}
\bibliography{refs.bib}

\newpage
\appendix

\begin{center}
    \LARGE\bf Appendices
\end{center}
\section{Approximate bounds for relationship between step length and time interval}
\label{app:steplength}

\subsection{No directional persistence}

In a random walk with no directional persistence, corresponding to a turning angle concentration equal to 0, \cite{kareiva1983} show that the squared displacement $R_n^2$ over $n$ time units satisfies
\begin{equation*}
    E[R_n^2] = n E[l^2]
\end{equation*}
where $l$ is the step length over a single time unit. For large enough $n$, \cite{bovet1988} further derived an approximate relationship between the expected displacement and expected squared displacement, which holds for large $n$,
\begin{equation*}
    E[R_n] \approx \frac{\sqrt{\pi}}{2} \sqrt{E[R_n^2]}.
\end{equation*}

Combining these two results, we get
\begin{equation*}
    E[R_n] \approx \frac{\sqrt{\pi E[l^2]}}{2} \sqrt{n}.
\end{equation*}

If $l \sim \text{gamma}(\mu_0, \sigma_0)$, where $\mu_0 > 0$ and $\sigma_0 > 0$ are the mean and standard deviation, respectively, it can be shown that $E[l^2] = \mu_0^2 + \sigma_0^2$. That is, the mean step length over $n$ time units is approximately
\begin{equation*}
    E[R_n] \approx \frac{\sqrt{\pi (\mu_0^2 + \sigma_0^2)}}{2} \sqrt{n}.
\end{equation*}

\subsection{Perfect directional persistence}

In a random walk with perfect directional persistence, the movement is in a straight line and the displacement after $n$ time units is simply the sum of $n$ step lengths: $R_n = l_1 + l_2 + \dots + l_n$. The expected displacement is therefore
\begin{equation*}
    E[R_n] = E[l_1 + \dots + l_n] = E[l_1] + \dots + E[l_n] = \mu_0 + \dots + \mu_0 = \mu_0 \cdot n
\end{equation*}

\section{$n$-step transition probabilities}
\label{app:tpm}

Consider a 2-state Markov chain with transition probability matrix
\begin{equation*}
    \bm\Gamma = 
    \begin{pmatrix}
    \gamma_{11}^{(0)} & \gamma_{12}^{(0)} \\
    \gamma_{21}^{(0)} & \gamma_{22}^{(0)}
    \end{pmatrix}
\end{equation*}
where $\gamma_{11}^{(0)} = 1 - \gamma_{12}^{(0)}$ and $\gamma_{22}^{(0)} = 1 - \gamma_{21}^{(0)}$. We want to find a general expression for the $n$-step transition probabilities in terms of the (one-step) transition probabilities in $\bm\Gamma$.

We can show that $\bm\Gamma = \bm{V} \bm{D} \bm{V}^{-1}$ where
\begin{equation*}
    \bm{V} =
    \begin{pmatrix}
        1 & \gamma_{12}^{(0)} \\
        1 & - \gamma_{21}^{(0)}
    \end{pmatrix}
    \quad
    \text{and}
    \quad
    \bm{D} =
    \begin{pmatrix}
        1 & 0 \\
        0 & 1 - \gamma_{12}^{(0)} - \gamma_{21}^{(0)}
    \end{pmatrix}
\end{equation*}

It is a well-known result that the $n$-step transition probability matrix is the $n$-th power of the one-step transition probability matrix, and therefore we seek
\begin{equation*}
    \bm\Gamma^{n} = (\bm{VDV}^{-1})^n = \bm{VDV}^{-1} \bm{VDV}^{-1} \bm{VDV}^{-1} \cdots \bm{VDV}^{-1} = \bm{V}\bm{D}^n\bm{V}^{-1}
\end{equation*}
where the $\bm{V}^{-1} \bm{V} = \bm{I}$ terms cancel out of the product. We have
\begin{equation*}
    \bm{D}^n =
    \begin{pmatrix}
        1 & 0 \\
        0 & (1 - \gamma_{12}^{(0)} - \gamma_{21}^{(0)})^n
    \end{pmatrix}
    \quad
    \text{and}
    \quad
    \bm{V}^{-1} = \frac{1}{\gamma_{12}^{(0)} + \gamma_{21}^{(0)}}
    \begin{pmatrix}
        \gamma_{21}^{(0)} & \gamma_{12}^{(0)} \\
        1 & - 1
    \end{pmatrix}
\end{equation*}

Combining these results, we find
\begin{equation*}
    \bm\Gamma^n =
    \frac{1}{\gamma_{12}^{(0)} + \gamma_{21}^{(0)}}
    \begin{pmatrix}
        \gamma_{21}^{(0)} + \gamma_{12}^{(0)} (1 - \gamma_{12}^{(0)} - \gamma_{21}^{(0)})^n & 
        \gamma_{12}^{(0)} [1 - (1 - \gamma_{12}^{(0)} - \gamma_{21}^{(0)})^n] \\
        \gamma_{21}^{(0)} [1 - (1 - \gamma_{12}^{(0)} - \gamma_{21}^{(0)})^n] & 
        \gamma_{12}^{(0)} + \gamma_{21}^{(0)} (1 - \gamma_{12}^{(0)} - \gamma_{21}^{(0)})^n
    \end{pmatrix}
\end{equation*}

\end{document}